\documentclass{article}

\usepackage{cite}
\usepackage{amsmath,amssymb,amsfonts}
\usepackage{algorithm}
\usepackage{algorithmic}
\usepackage{graphicx}
\usepackage{array}
\usepackage{textcomp}
\usepackage{xcolor}
\usepackage{fullpage}
\usepackage{caption}
\usepackage{subcaption}
\usepackage{float}
\usepackage{comment}
\usepackage{multirow}
\usepackage{footnote}
\usepackage{csquotes}
\usepackage{hyperref}
\usepackage{authblk}

\newcolumntype{P}[1]{>{\centering\arraybackslash}p{#1}}

\def\BibTeX{{\rm B\kern-.05em{\sc i\kern-.025em b}\kern-.08emhttps://www.overleaf.com/project/6166955754ac311fbee45991
    T\kern-.1667em\lower.7ex\hbox{E}\kern-.125emX}}

\begin{document}

\title{MERLIN - Malware Evasion with Reinforcement LearnINg}

\author[,1]{Tony Quertier\thanks{\texttt{tony.quertier@orange.com}}}
\author[1,2]{Benjamin Marais}
\author[1]{Stéphane Morucci}
\author[1]{Bertrand Fournel}
 
\affil[1]{Orange Innovation, Rennes, France}
\affil[2]{Department of Mathematics, LMNO, University of Caen Normandy, France}

\maketitle

\abstract{In addition to signature-based and heuristics-based detection techniques, machine learning (ML) is widely used to generalize to new, never-before-seen malicious software (malware). However, it has been demonstrated that ML models can be fooled by tricking the classifier into returning the incorrect label. These studies, for instance, usually rely on a prediction score that is fragile to gradient-based attacks. In the context of a more realistic situation where an attacker has very little information about the outputs of a malware detection engine, modest evasion rates are achieved \cite{Anderson2017}. In this paper, we propose a method using reinforcement learning with DQN and REINFORCE algorithms to challenge two state-of-the-art ML-based detection engines (MalConv \& EMBER) and a commercial antivirus (AV) classified by Gartner as a leader AV \cite{GarnerQuadrant2021EPP}. Our method combines several actions, modifying a Windows portable execution (PE) file without breaking its functionalities. Our method also identifies which actions perform better and compiles a detailed vulnerability report to help mitigate the evasion. We demonstrate that REINFORCE achieves very good evasion rates even on a commercial AV with limited available information.

}

%\keywords{malware evasion, reinforcement learning}

\section{Introduction}

Malicious software detection has become an important topic in business, as well as an important area of research due to the ever-increasing number of successful attacks using malware. According to Sophos \cite{SophosLtd2020} $73\%$ of ransomware attacks committed in 2020 were successful to encrypt computer files. AV-TEST \cite{avtest} estimates that about 450,000 new malware are discovered every day, and $93\%$ of them are Windows malicious files with a vast majority of portable executable (PE) files. Moreover, traditional signature-based methods cannot keep up with the rampant inflation of novel malware \cite{song2020mab}. Polymorphic malware use a mutation engine with self-propagating code to continually change its signature. As many anti-malware vendors use traditional signature-based detection methods to detect and block malicious code, it means that by the time they identify a new signature, the malware has already evolved into something new.

From an offense perspective, evading an antivirus (AV) is crucial for an attack to succeed. From a defense perspective, being assured that an AV is resilient to evasion techniques is essential. The goal of this paper is to address both challenges: making Windows PE malicious files undetected by AVs while providing explanations on why such evasion is successful. Identifying the root cause of evasions ( \textquote{explainability}) is a great value for AV technical designers when enhancing their heuristic detection modules.

We use a binary blackbox testing approach for attacking AVs according to coarse attacks classification suggested by Anderson and al in \cite{Anderson2017}: the system under attack is only required to report a malign or benign result for a given PE file. This makes an attack the most generic one, with low a priori knowledge requirements (neither the adversary model, the PE features space, nor the PE malware scores are required to conduct our attacks). Our system generates a modified version (an adversarial malware example) of a given PE file that leads to misclassification by AVs. By design, our semantic-preserving modifications do not corrupt the PE file substitute, i.e., the resulting evaded PE file keeps its malicious functionalities. 

In this work, we are focusing on static malware detection solutions since they are often the only security layers located on user devices and because such a static analysis is highly efficient to support large-scale analysis.  
This work shows that bypassing AVs can easily be automated with a sequence of clearly defined actions that may be of interest for AV designers when improving their detection engines.

\subsection{Background and related work}

Malware detection has become one of the top priorities of security actors since single incidences can cause millions of dollars worth of damages \cite{AndersonR2012}. The advances in the fields of artificial intelligence, ML, and deep learning make it possible to improve malware detection, and classification \cite{Raff2020, Ucci2019}. In particular, some notable datasets have been made publicly available, such as Ember \cite{Anderson2018}, SOREL-20M \cite{Sorel} or recently BODMAS \cite{bodmas}. These open datasets motivate new works, help in resolving existing challenges, and are very useful to benchmark new research proposals. ML-based classifiers can be designed and compared to keep track of the progress made by the research community. In \cite{Anderson2018}, Anderson et al. trained a feature-based malware detection model using a non-optimized LightGBM algorithm, whereas Raff et al. \cite{Raff2017} introduced MalConv, a featureless deep learning classifier using a dense neural network processing raw bytes of entire executable files. 

\begin{figure}[!ht]
    \centering
    \includegraphics[width=0.6\textwidth]{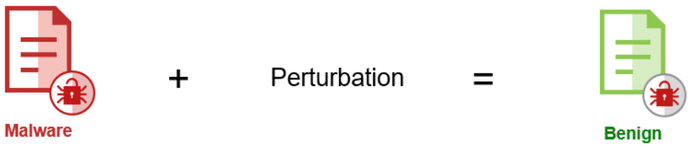}
    \caption{Malware adversarial example illustrated by Huang et al. \cite{Huang2019}}
    \label{fig:ae}
\end{figure}

Nevertheless, ML-based models may not be resilient to some attacks, namely \textquote{adversarial examples} which were firstly introduced by Szegedy et al. \cite{szegedy2013intriguing} and formalized by Goodfellow et al. \cite{goodfellow2014generative}. These examples are specifically crafted to perturb detection or classification models in a controlled way. The principle was initially described for image classification, but it was later generalized to other objects, and malicious files are not an exception. Figure \ref{fig:ae} illustrates how to generate an adversarial example by adding some perturbations to malware. Several approaches have been proposed to generate undetectable malicious files such as using DL techniques as GAN with MalGan \cite{Hu2017, Kawai2019} or MalFox  \cite{Zhong2020}. Other adversarial examples have already been described in literature \cite{Demetrio1, Kong2021, Park2020, Aryal2021, Demetrio2}. In particular, Demetrio et al. \cite{Demetrio1} use a particularly interesting method based on a constrained minimization problem while preserving executable files functionalities.

Anderson et al. \cite{Anderson2017} used reinforcement learning (RL) to automatically generate malicious files. They release on GitHub an RL framework ( \textquote{Gym-Malware}) that achieves an evasion rate of $16\%$ on 200 holdout samples when attacking a gradient boosted decision tree model trained on 100,000 malicious and benign samples. Song et al. \cite{song2020mab} propose an open-source RL framework called  \textquote{MAB-malware} also available on GitHub. Results show this framework can achieve a high evasion rate (over $75\%$) for two state-of-the-art ML malware detectors (EMBER \cite{Anderson2018}, and MalConv \cite{Raff2017}) and over $32\%$ to $48\%$ for commercial AVs in a pure black-box setting. Authors focus on re-implementing actions to transform malware without breaking its functionality since they found that more than $60\%$ of the transformed binaries with Gym-Malware were corrupted. Fang et al. \cite{Fang2020} propose a novel method for extracting PE file features and build two models called DeepDetectNet (defense) and RLAttackNet (attack), which contest with each other in the same way as Generative Adversarial Networks behave; they attack DeepDetectNet with RLAttackNet and achieve a success attack rate of $19.13\%$. They retrain DeepDetectNet with these new evaded malicious files, and the success attack rate dropped from $19.13\%$ to $3.1\%$. One of their conclusions is that ML-based models for malware detection can be improved using synthetically generated malware.  

\subsection{Contributions}

In this work, we first propose an RL framework to evade malicious file static-detectors based on state-of-the-art ML models. We begin by implementing a well-known RL model, Deep Q Network (DQN), and train it to evade malware on different models such as Malconv \cite{Raff2017}, LGBM Ember \cite{Anderson2018}, and Grayscale \cite{Marais2021}. The DQN model achieves very good results with Malconv and Grayscale with a respective evasion rate of $100\%$ and $98\%$. On Ember, its evasion rate reaches $67\%$, which motivated us to develop a better technique using the REINFORCE algorithm \cite{Willia1992}. To our knowledge, it is the first time such an algorithm has been used for malware evasion. We train REINFORCE against Ember, and our results show a slight improvement over DQN with an increase of the evasion rate from $67\%$ to $74.2\%$ without any impact on training time. We then challenge a well-known commercial AV. Once again, REINFORCE shows that it performs better than DQN with a significant increase of the evasion rate from $30\%$ to $70\%$.

A key element of our work is our ability to compile a vulnerability report listing the most efficient actions to transform a malicious PE file and make it undetectable by the model under attack. In other words, we can identify the detection model weaknesses and the most effective actions to defeat a given AV. Security experts can then leverage these insights to understand why a detection engine failed and react accordingly. 

Finally, our RL framework makes it possible to generate new malware variants and thus create a database of never-before-seen malicious files. This database could be used as a preventive asset to manage proactively potential malware variants.

\subsection{Outline}

In section 2, we introduce our RL approach. We describe the context, the agents, the
environments and the actions considered for evading an AV. In section \ref{sec:train_test}, we present the results of the training and testing phases. Next, in section \ref{sec:analysis_explainability}, we analyze the effectiveness of each agent and action for the different detection solutions by highlighting their weaknesses. Finally, section \ref{sec:conclusion} summarizes our results and describes future work.

\section{Algorithms and RL framework}
\label{sec:RL_approach}

\begin{figure}[!ht]
    \centering
    \includegraphics[width=0.7\textwidth]{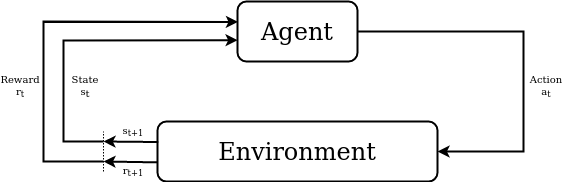}
    \caption{Reinforcement learning framework representation}
    \label{fig:RL}
\end{figure}

Our main objectives are (1) to bypass malware detection engines by generating undetected malicious PE files using an RL approach and (2) to provide an automated audit highlighting the weaknesses of a given AV. As shown in Figure \ref{fig:RL}, an RL framework consists of the interaction of an Agent with an Environment. At a given time $t$, the Agent observes the state $s$ of the Environment and performs an Action $a$ that modifies it. The Environment returns a Reward $r$, which depends on the effectiveness of this Action. In section \ref{subsec:agents}, we introduce the two agents that we have selected. Their job is to interact with a given PE file to modify it and make it undetectable by an ML solution or a commercial AV. Then, in section \ref{subSec:env}, we present environments that can output their states and rewards associated with two kinds of detection engines: ML models and a commercial AV. Finally, we detail some possible actions used to modify malicious files in section \ref{subsec:actions}. 

\subsection{RL agents}
\label{subsec:agents}

An agent is backed by different types of RL algorithms. Its function is to interact with an environment, modifying it in order to identify which actions perform best (see section \ref{subsec:actions}). Two RL models have been implemented at the agent side: Deep Q-Network (DQN) and REINFORCE. These two algorithms have their own characteristics and are presented in the following subsections.

\subsubsection{Deep Q-Network}

%DQN is probably the best-known model of deep RL. It is a Deep Q-Learning approach, firstly introduced by Mnih et al. \cite{Mnih2013}, based on Q-learning algorithm \cite{Watkins1992}.

The DQN algorithm, introduced by Mnih et al. \cite{Mnih2013}, is a popular deep RL model. Based on Q-learning algorithm \cite{Watkins1992}, its main advantage is that it can handle numerous and large states. The purpose of Q-learning algorithms and, by extension, Deep Q-learning algorithms is to predict the action that provides the best reward. For that, these two models approximate the optimal action-value function $Q^{*}(s,a)$. In the case of Deep Q-learning, the action-value function is modeled by a neural network called Q-Networks. % and is parametrized as $Q(s,a;\theta)$ where $\theta$ is the parameters of the action-values function and corresponding on the weights of the Q-networks. 
Based on the Q-learning update equation, presented in Equation  \ref{Eq:qlearning_update}, we update the weights of the Q-Network in an iterative way. After training our Q-network, our agent will return the best action $a$ associated with the best reward $r$, for a given state $s$. From a technical perspective, it is the best action to make our malicious file undetectable by the targeted AV.

\begin{equation}
\label{Eq:qlearning_update}
    Q(s_t,a_t) \leftarrow (1 - \alpha)Q(s_t,a_t) + \alpha(r + \gamma \max_{a}Q(s_{t+1},a))
\end{equation}

\subsubsection{REINFORCE}

%REINFORCE, also known as the Monte Carlo Policy Gradient method, is a policy gradient method. REINFORCE was firstly presented by Williams \cite{Willia1992} and it is a different RL approach from DQN that consists of optimizing a policy $\pi$. The approach maximizes the expected returned $J$ with respect to $\pi$ depending on parameters $\theta$. For this, it is necessary to apply the gradient descent method to $J(\theta)$ (Eq. \ref{Eq:REINFORCE1}). Gradient of $J(\theta)$, with respect to parameters $\theta$, is specified by Eq. \ref{Eq:REINFORCE_gradient} where $\tau$ is the trajectory, also known as episode, and consist on a sequence of states, actions and rewards indexed by $t\in[0, T]$. $Q$ is the action-value function of the policy $\pi_\theta$. So, for each state $s$, the REINFORCE agent returns a probability distribution, following the policy $\pi_\theta$, which is supposed to guide to the best-expected reward. figure \ref{fig:discounted_reward} illustrates this concept by comparing two different policies applied to the same problem.
 
REINFORCE, also known as the Monte Carlo Policy Gradient method, was introduced by Williams in \cite{Willia1992}. It is another RL approach that consists in maximizing the expected returned $J$ with respect to the policy $\pi$, depending on parameters $\theta$. For this, it is necessary to apply the gradient descent method to the function  

\begin{equation}
    J(\theta) = \mathbb{E}_{\tau \sim \pi_{\theta}}[\sum_{t=0}^{T} Q_{\pi_{\theta}}(s_{t}, a_{t})].
\end{equation} 

% E = Esperance ? 
% J = Jacobian ? 

\begin{figure}[!ht]
    \centering
    \includegraphics[width=1\textwidth]{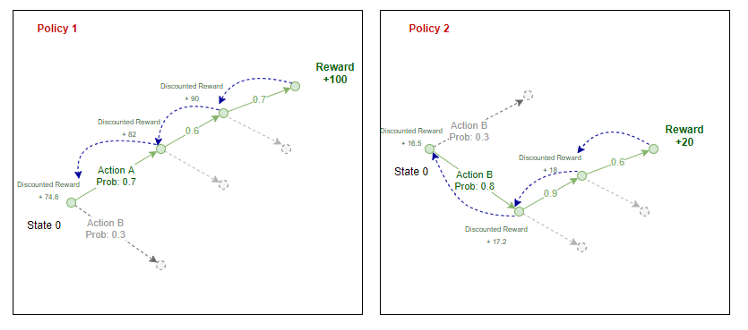}
    \caption{Two different policies \cite{web1}}
    \label{fig:discounted_reward}
\end{figure}

The gradient of $J(\theta)$, with respect to parameters $\theta$, is specified by 
\begin{equation}
    \nabla J(\theta) = \mathbb{E}_{\tau \sim \pi_{\theta}}[\sum_{t=0}^{T} \nabla_{\theta}\log \pi_{\theta}(a_{t}|s_{t})Q_{\pi_{\theta}}(s_{t}, a_{t})],
\end{equation}
where $\tau$ is the trajectory, also known as episode. The trajectory consists of a sequence of states, actions and rewards indexed by $t\in[0, T]$, and $Q$ is the action-value function of the policy $\pi_\theta$. So, for each state $s$, the REINFORCE agent returns a probability distribution, following the policy $\pi_\theta$, which is expected to lead to the best expected reward. Figure \ref{fig:discounted_reward} illustrates this concept by comparing two different policies applied to the same problem.

\subsection{Environments}
\label{subSec:env}

ML detection solutions are a convenient way to formalize our problem because they provide us with a lot of useful information. For each of them, we have the features used to detect whether the scanned file is malicious or not. These models also return a prediction score, or detection score, $p \in [0,1]$ where the closer to $1$, the more likely it is that the file is malicious. This score provides important information during the RL process. We analyse two state-of-the-start detection engines (Ember \cite{Anderson2018} and Malconv, two references in the research community, \cite{Raff2017}) together with our own model  \textquote{Grayscale} \cite{Marais2021}.

We also select a commercial AV since we hope that our work could help designers improve their detection capabilities. Unlike ML models, commercial AVs are blackboxes and they usually return very little information after a scan: a hard detection label $l \in \{0,1\}$, $0$ if the file is benign or $1$ if it is malicious.

As a consequence, our environments must take into consideration the quantity of the available information when using a detection tool. Table \ref{tab:knowledge} summarizes this information.

\begin{table}[!ht]
\begin{center}
\caption{Knowledge available for each detection tool}
\label{tab:knowledge}
    \begin{tabular}{|P{2.8cm}|P{1.5cm}|P{1.5cm}|P{1.5cm}||P{2.8cm}|}
    \hline
    \multicolumn{1}{|c|}{} &
    \multicolumn{3}{c||}{ML solutions} &
    \multicolumn{1}{c|}{ Commercial AV} \\
    \hline
    Models & Ember & Malconv  & Grayscale &  \\
    \hline
    Original file & X & X & X & X \\ \hline
    Extracted features & X & X & X & \\ \hline
    Prediction score & X & X & X &  \\ \hline
    Detection label & X & X & X & X \\ 
    \hline
    \end{tabular}
\end{center}
\end{table}

\begin{figure}[!ht]
    \centering
    \includegraphics[width=0.7\textwidth]{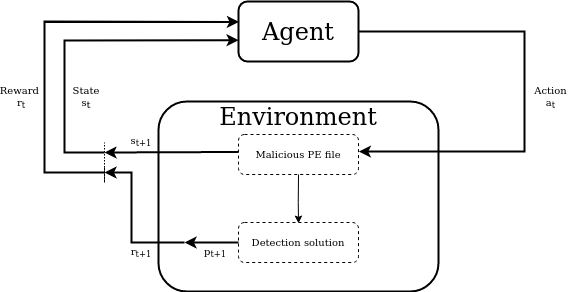}
    \caption{Reinforcement learning framework representation with detailed environment}
    \label{fig:RL2}
\end{figure}

\subsubsection{ML solution}

\label{Sec:evading_ml}

The first step is to design some environments with which the agents can interact. These environments must be able to:

\begin{enumerate}
    \item generate a new episode i.e. pick a new malicious file to work on; 
    \item apply a number of modifications (i.e. the actions $a_t$ selected by our agents) on the malicious file;
    \item for any modification, return:
    \begin{itemize}
         \item[$\bullet$] a reward $r_{t+1}$ depending on the effectiveness of the action,
         \item[$\bullet$] some features, extracted from the new PE file, corresponding to the state $s_{t+1}$.
    \end{itemize}
\end{enumerate}

Table \ref{tab:inputShape} summarizes the different characteristics of the detection models. To make our approach generic, we create a dedicated environment for each one. As shown in figure \ref{fig:RL2}, each environment contains a list of malicious PE files required for the training and testing phases and the detection model that the agent must bypass.

%We create a dedicated environment for each detection model. Each model has its own characteristics summarized in table \ref{tab:inputShape}. And so, as shown in figure \ref{fig:RL2}, an environment is composed of a list of malicious PE files required for training and testing the RL algorithms and the detection model that agents must disrupt. 

%figure \ref{fig:RL2} describes in detail how our environment is built. For each malware detection model, we create a specific environment. Indeed, each model is based on a certain ML algorithm and uses a different data format than the others. table \ref{tab:inputShape} summarizes the information about each ML detection model.

We also need to define the reward function were the more effective an action, the greater the reward. This function is given below:

\begin{comment}
\begin{equation}
\label{eq:r1}
    \text{for each iteration $t \in[1,T]$ : } 
    r_t = r(p_{t}, p_{t-1}) = 
    \begin{cases}
    R \in \mathbb{R} & \quad \text{if } p_{t} < \phi \\
    p_{t} - p_{t-1} & \quad \text{else}
    \end{cases}
\end{equation}
\end{comment}

For each iteration $t \in[0,T]$, we set
\begin{equation}
\label{eq:r1}
 r_t := r(p_{t}, p_{t-1}) = 
\begin{cases}
    R \in \mathbb{R} & \quad \text{if } p_{t} < \phi \\
    p_{t} - p_{t-1} & \quad \text{otherwise}
\end{cases}
\end{equation}

where $p_{t}$ is the detection score returned by the detection model, $\phi \in \mathbb{R}$ is a threshold depending on the model (see Table \ref{tab:inputShape}) and $R$ is an hyperparameter chosen according to our tests.

\begin{savenotes}
\begin{table}[!ht]
\begin{center}
\caption{ML detection models summary}
\label{tab:inputShape}
    \begin{tabular}{|P{2.5cm}|P{2.7cm}|P{2.7cm}|P{2.7cm}|}
    \hline
    \centering
        & Ember & Malconv & Grayscale \\ 
        \hline
        Type of model & LGBM & DNN & CNN \\
        \hline
        Type of data & Features extract from PE files \cite{Anderson2018} & Raw bytes extract from PE files \cite{Raff2017} & PE files converted into images \cite{Nataraj2011} \\
        \hline
        Size of state\footnote[1]{ corresponding to the size of the features vector extracted from PE file}
        & 2381 & $\approx$1M & $64\times64$ \\
        \hline
        Threshold $\phi$ & 0.8336 & 0.50 & 0.50 \\
    \hline
    \end{tabular}
\end{center}
\end{table}
\end{savenotes}

\subsubsection{Commercial AV}
\label{Subsec:evading_av}

The main challenge in creating a suitable environment for the commercial AV is related to a lack of information. We only have at our disposal the PE malicious file, which does not provide any proper observable space, i.e., a state $s$. To get around this limitation, the missing state $s$ is replaced by some features of the PE file. These characteristics are returned by Ember features extractor, but any other solution could also be used if necessary. The key here is to have a features vector that models states to apply RL algorithms on the AV.  

Another problem is that the commercial AV does not return a score $p \in [0,1]$, as for ML solutions, but a hard label $l \in \{0,1\}$. As a consequence, the reward function in equation \ref{eq:r1} does not apply. We need to define a new reward function dedicated to commercial AV as follows:

for each iteration $t \in[0,T]$, we set
\begin{equation}
\label{eq:r2}
r_t := r(l_t) = 
    \begin{cases}
    R \in \mathbb{R} & \quad \text{if } l_t = 0 \\
    0 & \quad \text{otherwise}
    \end{cases}
\end{equation}
where $l_{t}$ is the label returned by commercial AV.

\begin{comment}
Now that the main limitations related to a commercial AV are no longer a problem: the two RL agents presented earlier can handle this new environment. 
\end{comment}

\subsection{Actions}
\label{subsec:actions}

In this section, we are listing the actions that are applied to modify a given PE file, with the constraint of not corrupting its functionalities. Among other things, we use the actions introduced by Anderson et al. \cite{Anderson2017}. They are listed in Table \ref{tab:action_table}. In particular, these actions include:

\begin{itemize}
 \item[$\bullet$] packing and unpacking the malicious file,
 \item[$\bullet$] adding benign strings to the end of sections,
 \item[$\bullet$] adding random bytes to the unused section space,
 \item[$\bullet$] adding import functions,
 \item[$\bullet$] modifying PE files timestamp.
\end{itemize}

Because PE file modifications must be coherent and legitimate, we have previously constructed a database containing relevant information about 5000 benign files (section names, imports, section strings, etc.). 
When specific content is required by an action (like an import, the adding section or strings, etc.), it is randomly picked from this database. While testing, we refined and reduced the size of this database, keeping only the elements that were the most efficient in order to limit the action space of our agent.

\begin{table}[]
    \caption{Action table. Actions are 0-indexed}
    \label{tab:action_table}
    \centering
    \begin{tabular}{|c|P{1cm}|}
        \hline
        Action & Id \\ 
        \hline
        modify machine type & 0 \\
        pad overlay & 1 \\
        append benign data overlay & 2 \\ 
        append benign binary overlay & 3 \\ 
        add bytes to section cave & 4 \\ 
        add section strings & 5 \\ 
        add section benign data & 6 \\ 
        add strings to overlay & 7 \\ 
        add imports & 8\\
        rename section & 9\\
        remove debug & 10\\
        modify optional header & 11\\
        modify timestamp & 12\\
        break optional header checksum & 13\\
        upx unpack & 14\\
        upx pack & 15\\
    \hline
    \end{tabular}
\end{table}

\begin{figure}[h!]
    \centering
    \includegraphics[width=0.9\textwidth]{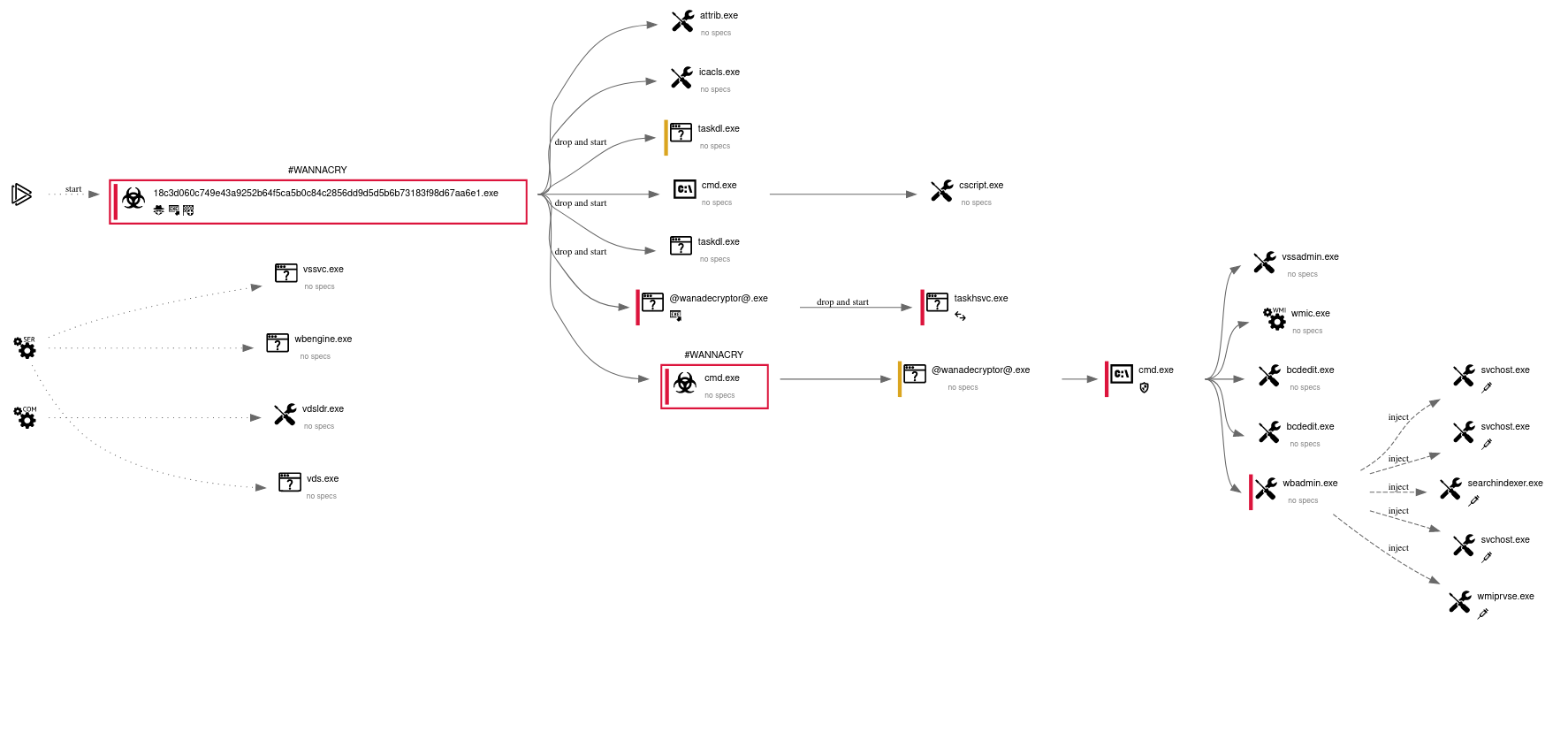}
    \caption{Behavior graph of the modified Wannacry malware from malware sandbox ANY.RUN}
    \label{fig:Wannacry}
\end{figure}

LIEF \cite{LIEF} is used to apply actions and make modifications on a PE file, but, as mentioned by Song et al. in \cite{song2020mab}, this library may sometimes break PE functionalities. To circumvent this problem, Song et al. analyzed their samples in the Cuckoo sandbox. 
During our RL process, we follow a similar method: if the file is broken after some modifications performed by the agent, we mark the evasion as failed and start again. When an evasion is successful, we validate that the malware functionalities have not changed by checking with an off-the-shelf sandbox and with an internal custom-built one where all system calls are logged, making it possible to compare execution flows of both modified and original files. We also verify on \href{https://any.run}{ANY.RUN} that the behavior graphs of the modified and the original are similar. Such a behavior graph is presented in figure \ref{fig:Wannacry} and shows the graph of a modified Wannacry malware.

In the next section, we detail our results with a particular focus on the effectiveness of each action during the training and testing phases. 

\section{Experiments}
\label{sec:train_test}

During our first experiments, we noticed that malware from different open datasets were not equal from an evasion perspective: some were quite simple to evade, whereas others from the BODMAS Malware Dataset \cite{bodmas} (created and maintained by Blue Hexagon and UIUC) led to some evasion difficulties. We suppose that these differences may be related to the age of the malware: the BODMAS dataset is quite recent as it contains 57,293 malware samples collected from August 2019 to September 2020. In this paper, we decided to rely only on this dataset for malware selection.

\begin{comment}
The agents need to be trained to learn how to bypass the different detection solutions.  
\end{comment}

\begin{comment}
We first experimented with malware from various open datasets. Most of the malware was too simple to evade, so there is not much interest in using such complex learning methods. We, therefore, decided to use the BODMAS Malware Dataset \cite{bodmas} created and maintained by Blue Hexagon and UIUC. The BODMAS dataset contains 57,293 malware samples collected from August 2019 to September 2020.
\end{comment}

As mentioned in the previous section, the results provided by ML solutions and a commercial AV are quite different (a scoring label vs. a hard label). Moreover, our commercial AV has significantly higher response times than ML solutions. As a consequence, we separate our experiments: for ML solutions (Malconv, Grayscale, and Ember), we use 1,000 malicious files for training and 500 malicious files for testing. For the commercial AV experiment and mainly due to its response times, we use only 100 malware for training and 50 malware for testing.

\begin{comment}
The experiments are separated according to whether we are targeting an ML detection solution or a commercial antivirus.   Furthermore, due to processing time (we had limited access to the antivirus API), we use 1,000 malicious files from BODMAS to train our agents on Malconv, Grayscale, and Ember, and 500 to test them. For the antivirus experimentation, we only use 100 malware to train it and 50 to test it because processing time 
\end{comment}

\begin{comment}
In this work, we did not train and test REINFORCE on Malconv and Grayscale since they both are trivial to evade using the DQN agent, as demonstrated in the following paragraphs. We limit our experiments with REINFORCE on two AV solutions that are more difficult to bypass: Ember and a commercial AV.
\end{comment}

\subsection{Malconv Evasion}
\label{malconv}

\begin{figure}[!ht]
    \centering
    \includegraphics[width=0.8\textwidth]{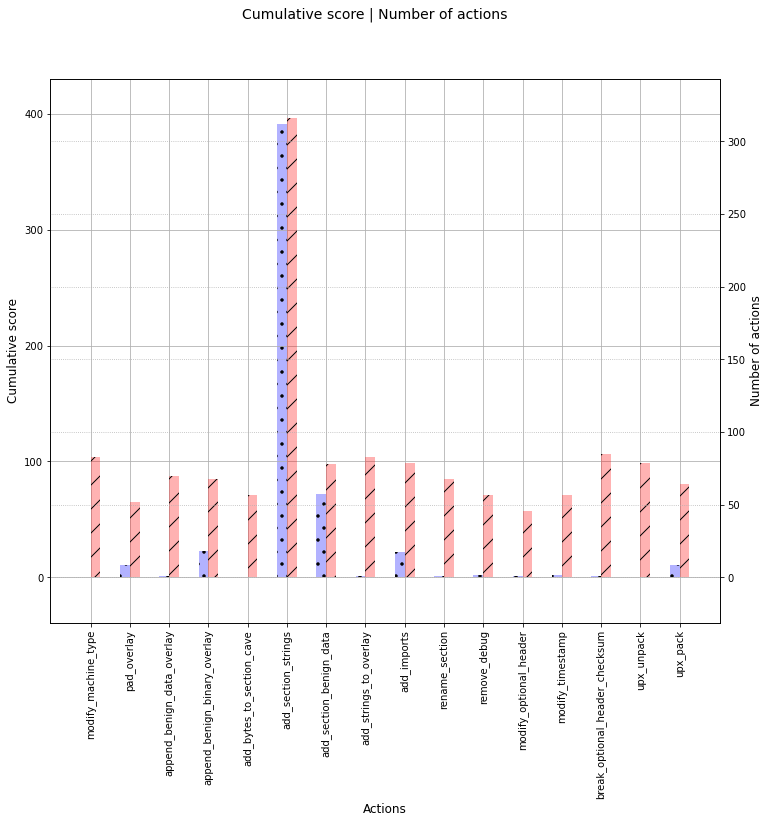}
    \caption{Cumulative score (left) and number of action (right) \\
    Training of DQN agent on Malconv}
    \label{fig:malconv_train}
\end{figure}

We begin by training the DQN agent to evade Malconv. Results in figure \ref{fig:malconv_train} display the resulting cumulative scores when using a given action (left hatched bar) and the number of times such an action has been applied (right dotted bar). Since we rely thoroughly on this kind of visualization in the rest of this article, we are detailing hereafter how it can be interpreted. In figure \ref{fig:malconv_train}, the action \textquote{modify machine type}, indexed by $0$, is used about $100$ times and it does not generate any score. This information is useful in itself even if the score is null because it shows that this action is useless during training wherever it has been applied (we are combining actions into a sequence to evade an AV and the positions in this sequence where this action is applied may influence the evasion score differently). The so-called  \textquote{add section strings} action, indexed by $5$, is the most used and produces a high cumulative score, which means that this action is really efficient to bypass Malconv. We can also notice that action 6 (\textquote{add section benign data}) has an interesting ratio between its number of usage and its cumulative score. Since Malconv is a Natural Language Processing (NLP)-based DL algorithm, it is, therefore, no surprise that Malconv is not very robust to strings injection. Our agent reaches an evasion rate of $96\%$ against Malconv by identifying its main weakness (being the \textquote{add section strings}). 

During the testing phase, the agent only uses action 5 to achieve an evasion rate of $100\%$. We do not discuss the Malconv detection algorithm any further since this result shows that Malconv is easy to evade with just one action, and some articles have already detailed its strengths and weaknesses as in \cite{fleshman2018non, Kolosnjaji2018, Demetrio3} for example.

%We begin by training the DQN agent to fool Malconv. As we can see in figure \ref{fig:malconv_train}, the so-called "add section strings" action (action 5, actions being 0-indexed) is the most used, and its cumulative score is very high, meaning that it is very efficient to bypass Malconv. We can also notice that action 6 ("add section benign data") has an interesting ratio between use and score. Since Mavconv is a Natural Language Processing (NLP)-based DL, it is, therefore, no surprise that Malconv is not very robust to strings injection.
%The results are consistent, as Malconv is a deep learning algorithm based on NLP (Natural Language Processing), adding strings strongly modifies the scoring. 

\subsection{Grayscale Evasion}
\label{grayscale}

\begin{figure}[!h]
    \centering
    \includegraphics[width=0.8\textwidth]{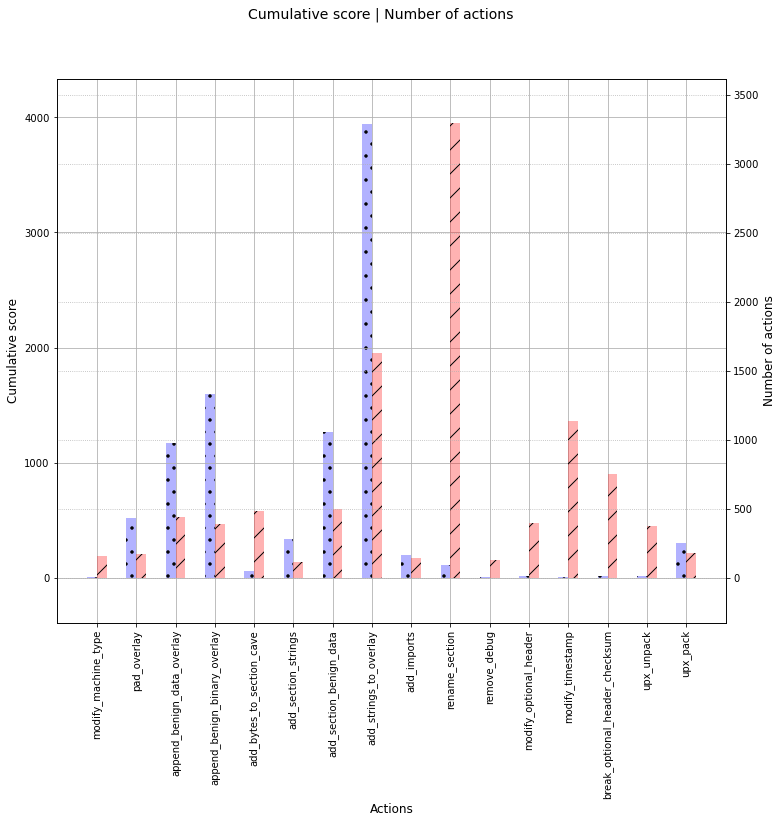}
    \caption{Cumulative score (left) and number of action (right) \\
    Training of DQN agent on Grayscale}
    \label{fig:grayscale_train}
\end{figure}

Grayscale detection algorithm \cite{Marais2021} relies on transforming binary PE files into grayscale images and detecting malware using a trained CNN (Convolutional Neural Network) model. The evasion rate during the learning phase is $90 \%$. Unlike Malconv where the agent overfits by relying on just one action, this grayscale-based agent uses a combination of actions. The average number of steps to evade malware is about $10$ for the $1,000$ malware samples. Moreover, as we can see in figure \ref{fig:grayscale_train}, several hatched bars lead to positive cumulative scores, meaning that the associated actions of evading the AV are quite diversified (actions \{2,3,6,7\}), which is an advantage as suggested later in this paper. During the testing phase, the agent achieves an evasion rate of $98\%$ with the actions identified during the training phase.

As with Malconv, the Grayscale model is easy to bypass mainly due to the fact that it uses images representation of malicious files, which can be assimilated to file-signatures.

\subsection{Ember LGBM Evasion}
\label{ember}

Ember \cite{Anderson2018} is a LGBM model trained on the Ember dataset which contains one million malicious files. Due to this large sample size, we assumed that Ember would be harder to evade. We first trained the DQN agent against Ember and then against the REINFORCE agent.

\subsubsection{DQN algorithm}

\begin{figure}[h!]
    \centering
    \includegraphics[width=0.8\textwidth]{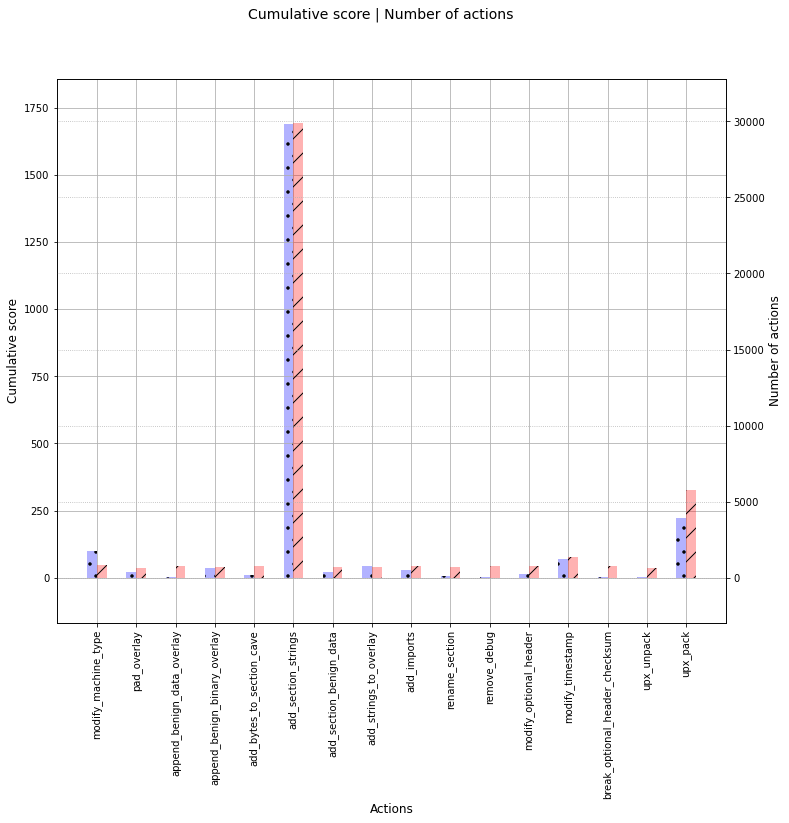}
    \caption{Cumulative score (left) and number of action (right) \\
    Training of DQN agent on Ember}
    \label{fig:ember_train}
\end{figure}

The DQN agent reaches an evasion rate of $40\%$ during the training phase. As assumed, Ember is more robust when compared to Malconv or Grayscale. Moreover, the average number of steps to make a malicious file undetectable is higher and reaches $30$, versus only $10$ on average for Malconv. This shows that the DQN agent has some difficulties in learning what the best required actions to evade are. In figure \ref{fig:ember_train} we notice that action 5 is the action that produces the best cumulative score. To a lesser degree, some actions like, for example, \textquote{upx pack}, indexed by $15$, have some impact and may be considered. Since this action has a higher relative impact than on Malconv and Grayscale, one could infer that their detection techniques are different.

These results are confirmed during the testing phase. The agent only uses action 5 until the malware is no longer detected by Ember. As for Malconv, DQN tends to overfit by relying on just one action. It reaches an evasion rate of $67\%$ for an average number of steps of $8$. Results are better than during the training phase because the agent has already learned and found one Ember weakness ( \textquote{add section strings}). Nevertheless, results are not as good as for Malconv and Grayscale.  

\subsubsection{Reinforce algorithm}

\begin{figure}[!h]
    \centering
    \includegraphics[width=0.8\textwidth]{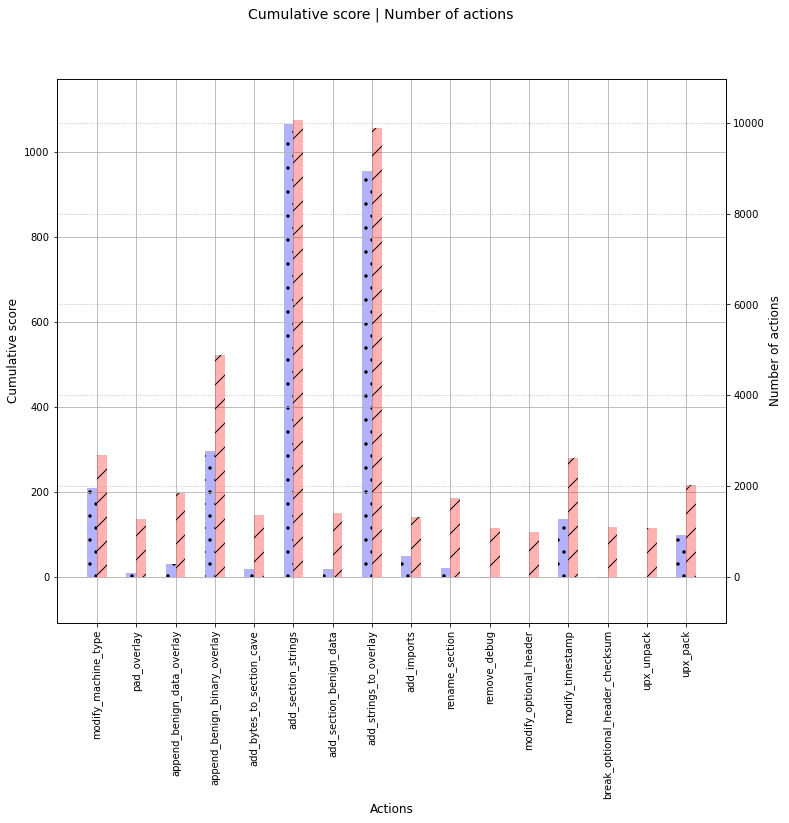}
    \caption{Cumulative score (left) and number of action (right) \\Training of REINFORCE agent on Ember}
    \label{fig:ember_reinforce_train}
\end{figure}

\begin{comment}
To prevent the agent from focusing on a single action and also to have a better performance than the DQN agent, we implement and train the REINFORCE agent.
\end{comment}
During training, the REINFORCE agent reaches an evasion rate of $54.7\%$ with $8$ steps on average. At first sight, the REINFORCE agent performs better than the DQN agent (DQN evasion rate is $40\%$). As shown in figure \ref{fig:ember_reinforce_train}, actions \{5, 7\} are largely dominant, but we can see that effective actions are more diversified, and their cumulative scores are better than during the training phase of the DQN agent. As with Malconv, adding strings in the binary file is an effective technique to perturb Ember. This lack of robustness is consistent with the study of Oyama et al. \cite{oyama2019identifying} which explains that only a few features contribute to explain Ember's results.

\begin{figure}[!h]
    \centering
    \includegraphics[width=0.8\textwidth]{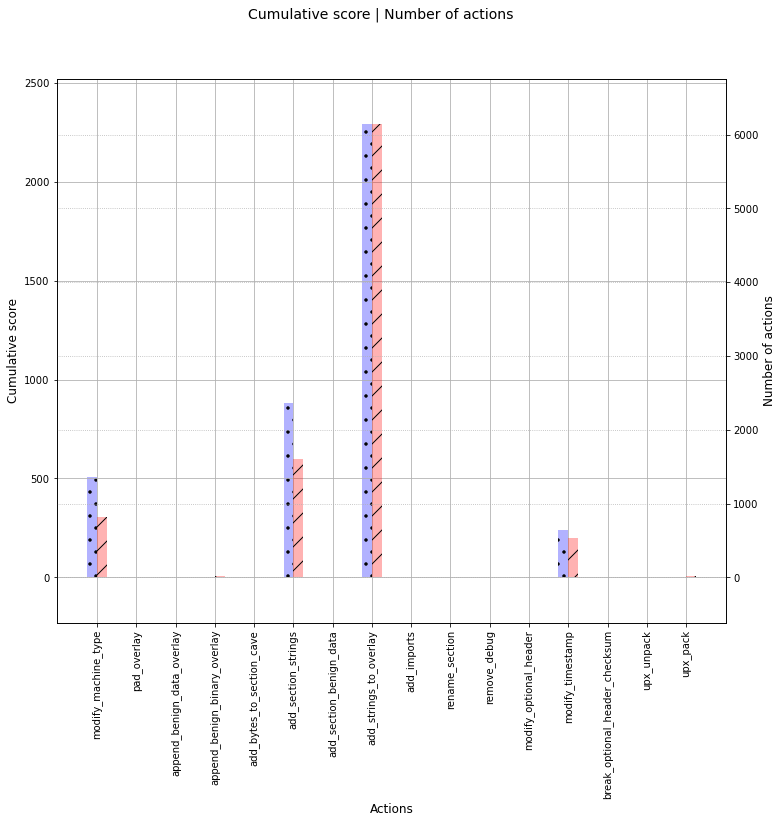}
    \caption{Cumulative score (left) and number of actions (right) \\ Testing of REINFORCE agent on Ember}
    \label{fig:ember_reinforce_test}
\end{figure}

Test results are also better with the REINFORCE agent than with the DQN agent, with a slight increase of the evasion rate from $67\%$ (DQN) to $80 \%$ (REINFORCE) and an average number of steps of $7.5$ (8 for DQN). We also observe in figure \ref{fig:ember_reinforce_test} that not only action 5 perturbs Ember, but also actions \{0, 7, 12\}.

\subsection{Commercial AV}
\label{antivirus}

As explained in section \ref{subSec:env}, an important limitation of our commercial AV is that it only returns a hard label (0 or 1) when scanning a file. Therefore, when our agents try to bypass it, the real impact and contribution of each action are much more difficult to assess than with an ML model. The following sections present our results with a commercial AV when training the DQN and the REINFORCE agents.  

\subsubsection{DQN algorithm}

\begin{figure}[h!]
    \centering
    \includegraphics[width=0.8\textwidth]{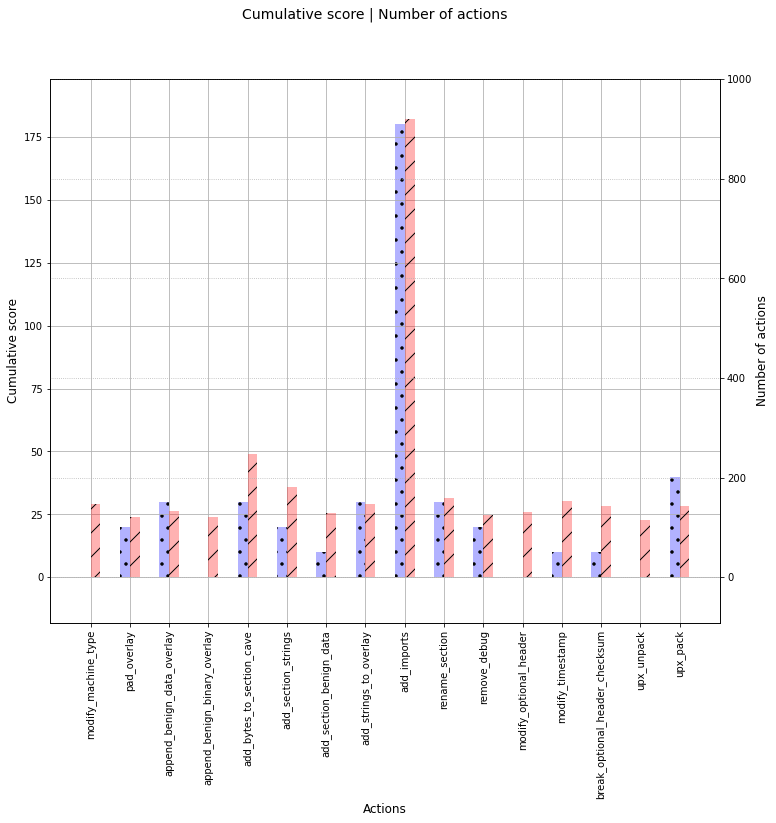}
    \caption{Cumulative score (left) and number of action (right) \\
    Training of DQN agent on AV}
    \label{fig:McAfee_dqn_train}
\end{figure}

The DQN agent reaches an evasion rate of $61\%$ in $9$ steps, on average, during the training phase. Nevertheless, as for Ember, we observe in figure \ref{fig:McAfee_dqn_train} that action 8 is dominant and returns the best cumulative score, even if some other actions are worth considering since they output non-zero rewards. As a reminder, for each episode, only the last action of the sequence gets a non-zero score due to the hard labeling of the commercial AV. 
%So, we can see that several different actions have allowed evading malware as several actions have a non-zero cumulative score in figure \ref{fig:McAfee_dqn_train}.

As for Ember, during the testing phase, only the action that performed best during the training phase is used. But unlike Ember, the DQN agent does not perform well with an evasion rate of $30\%$. This poor performance is due to the fact that we no longer have a score when scanning a modified PE file. The algorithm learned that only action 8 is useful to evade malware, but this result can be erroneous: let us imagine that the action applied before action 8 did most of the evasion job and that the last action (action 8) just terminates the RL job with a very limited impact on the internal (and unavailable) commercial AV prediction score. In this situation, only the last action will get the reward instead of the previous legitimate action. With Ember, the algorithm showed that a single action contributed greatly to the drop in detection score, but this is not the case here with the commercial AV. The low evasion rate is due to the fact that DQN agent focuses only on the action that evades the malware (because we have a label and not a score as with Ember) and not on those before. 

\subsubsection{REINFORCE algorithm}

In section \ref{ember} we show that the REINFORCE agent reaches a high evasion rate using more diversified actions. Unlike DQN, it has the advantage of not focusing on a single action. With a commercial AV, the REINFORCE algorithm reaches an evasion rate of $63\%$ during training, which is close to the results of the DQN agent ($61\%$). Also, the most impactful action is action 8, demonstrating that adding new imports to a malicious file is the best way to bypass this commercial AV.

\begin{figure}[h!]
    \centering
    \includegraphics[width=0.8\textwidth]{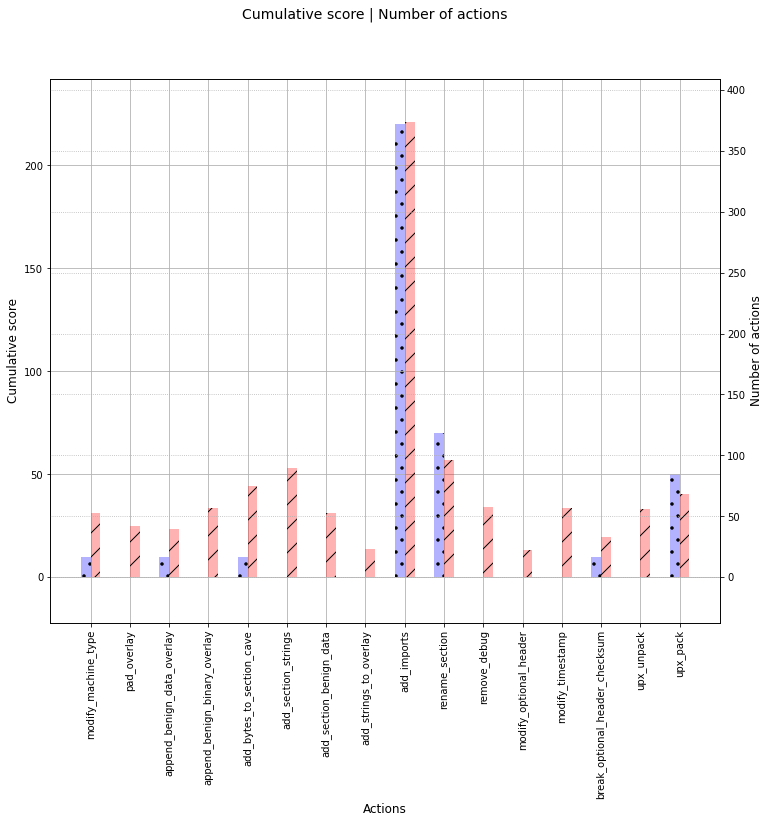}
    \caption{Cumulative score (left) and number of action (right) \\
    Training of REINFORCE agent on AV}
    \label{fig:McAfee_reinforce_train}
\end{figure}

During the test, the REINFORCE agent uses not only action 8 but also some other actions that have positive cumulative scores. During the testing phase, the agent reaches an evasion rate of $70\%$ for an average number of steps of $8$. As in the case of Ember, the REINFORCE agent performs better than the DQN agent in terms of evasion success. 
\begin{comment}
It is interesting to use the REINFORCE algorithm, which learns a sequence of actions, as an alternative solution when we only have a hard label and not a score.
Ah bon pourquoi ?
\end{comment}

\begin{figure}[h!]
    \centering
    \includegraphics[width=0.8\textwidth]{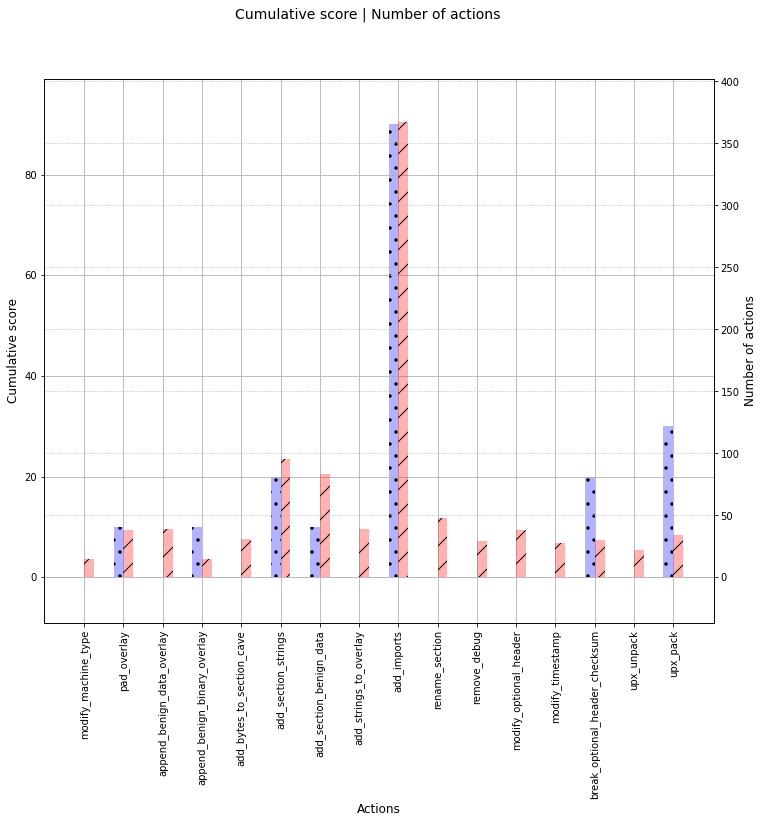}
    \caption{Cumulative score (left) and number of action (right) \\
    Testing of REINFORCE agent on AV}
    \label{fig:McAfee_reinforce_test}
\end{figure}

We argue that the REINFORCE algorithm outperforms other algorithms and that the identification of diversified actions is an additional and strong advantage of this agent.

\section{Analysis and explainability of results}
\label{sec:analysis_explainability}

In this section, we firstly discuss and summarize the results of the experiments. Then, we present our so-called  \textquote{vulnerability report} which is a  complementary tool to the charts presented in section \ref{sec:train_test}. Indeed, besides identifying which actions perform better, we can analyze what they actually did, what are the relevant imports for each algorithm, which strings have the most influence on the detection score, etc. We can also visualize which parts of the binary have been modified by a given action by transforming the malware into an image. Such insights are key elements when improving malware detection engines. Finally, we provide examples of malware evasion with concrete cases. 

%In this section, we discuss the results of our experiments. We are convinced that a great advantage of our work is that we provide visualization tools that support the understanding of the results. Besides finding out which actions are the most effective, we can analyze what each action actually did, what are the effective imports for each algorithm, which strings have the most influence on the detection score, etc. We can also visualize which parts of the binary have been modified by a given action by transforming the malware binary into an image.
%It is very important in order to improve malware detection tools, to understand where their weaknesses come from. 

\subsection{Summary of the results} %Je ne suis pas convaincu par le titre, à voir

Table \ref{tab:er} summarises the evasion rates of each agent for each detection solution during the tests. We also add a random agent, who selects actions randomly according to a uniform distribution, to compare the results. 
The REINFORCE agent outperforms every other agent on Ember and a commercial AV. %Furthermore, the efficiency of REINFORCE is better than the random agent. 
%We can conclude that REINFORCE learned better during the training on how to bypass the different detection solutions.

Table \ref{tab:abg_step} presents the average number of steps to evade a malicious file. Except for the commercial AV, REINFORCE is slightly more efficient. Nevertheless, it takes only $3$ more steps on average for a major increase in the evasion rate. 

\begin{comment}
We also assume that it has learned to apply some action sequences, in particular against the commercial AV.
\end{comment}

One possible explanation for the efficiency of REINFORCE is that it learns sequences of actions and not just the best action at a given time $t$. As a reminder, REINFORCE uses a policy to determine which actions are most appropriate to successfully bypass a targeted model, as illustrated in figure \ref{fig:discounted_reward}. Indeed, as we can see in the figure \ref{fig:McAfee_reinforce_test}, the agent uses different actions that do not provide a positive reward. Nevertheless, even if the last action of each episode gets a positive reward when the evasion is successful, the other actions appear to be useful to achieve this goal.

\begin{table}[!ht]
\begin{center}
\caption{Evasion rate of each algorithm during testing phase}
\label{tab:er}
    \begin{tabular}{|P{2.8cm}|P{1.5cm}|P{1.5cm}|P{1.5cm}||P{2.8cm}|}
    \hline
    \multicolumn{1}{|c|}{} &
    \multicolumn{3}{c||}{ML solutions} & 
    \multicolumn{1}{c|}{ Commercial AV} \\\cline{2-5}
    & Ember & Malconv  & Grayscale & commercial AV \\
    \hline
    DQN & $67 \%$ & $100 \%$ & $98 \%$ & $30 \%$ \\ \hline
    REINFORCE & $80 \%$ & $100 \%$ & $100 \%$ & $70 \%$ \\ \hline
    Random & $25 \%$ & $76 \%$ & $90 \%$ & $50 \%$ \\ \hline
    \end{tabular}
\end{center}
\end{table}

\begin{table}[!ht]
\begin{center}
\caption{Average number of step to evade malicious files for each agent}
\label{tab:abg_step}
    \begin{tabular}{|P{2.8cm}|P{1.5cm}|P{1.5cm}|P{1.5cm}||P{2.8cm}|}
    \hline
    \multicolumn{1}{|c|}{} &
    \multicolumn{3}{c||}{ML solutions} & 
    \multicolumn{1}{c|}{ Commercial AV} \\\cline{2-5}
    & Ember & Malconv  & Grayscale & commercial AV \\
    \hline
    DQN & $8$ & $9.5$ & $10$ & $5$ \\ \hline
    REINFORCE & $7.5$ & $8$ & $9$ & $8$ \\ \hline
    Random & $7$ & $20$ & $14$ & $12$ \\ \hline
    \end{tabular}
\end{center}
\end{table}

\subsection{Explainability method}

Explainability is a key advantage when dealing with ML results, especially in critical fields of application like Cybersecurity, for instance. 
%To go one step further, we decided to add a component to our RL framework to help explain the results. 
In addition to bypassing the detection solution, we introduced an additional component to record all modifications done to a PE file to make it undetectable. The purpose is to understand (explain) how an action transforms the binary and why this change allows bypassing the detection solution.

Some actions add information to the binary file like actions \{5, 6, 7, 8\}, while others modify some parts of the file like actions \{9, 11, 12\}. Since action 8 \textquote{add imports} for example, randomly picks a library in a predefined set, it does not generate the same reward. By recording the different impact events related to each distinct import, it is now possible to determine which imports are the more effective. This kind of information is very important for security analysts to help them improve their detection engines and locate potential weaknesses.
%By looking deeper at the different imports used, it allows determining which imports are more efficient to bypass a malware detection tool. For a malware analyst or security software company, this could be a preventive solution to reinforce detection solutions and a tool to locate any potential weaknesses.

\begin{figure}[h!]
    \centering
    \includegraphics[width=0.7\textwidth]{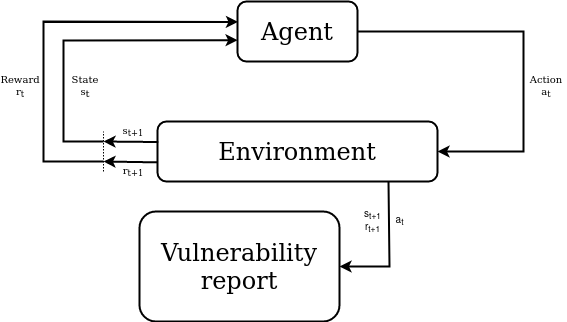}
    \caption{Reinforcement learning framework with the  \textquote{vulnerability report} component}
    \label{fig:RL3}
\end{figure}

Figure \ref{fig:RL3} illustrates how we store the relevant information, i.e., the actions, the states, and the rewards, after modifying the PE file in a vulnerability report. Any other useful information is also recorded in this report. For example, if the agent applies action 8, we add to the report the specific import that has been selected.

This report allows us to keep track of all changes that make malware undetectable. Also, it complements the information provided by all our visualization tools about the effectiveness of an action. The visualization tool and the report could give cybersecurity experts a turnkey solution to enhance detection solutions by fixing their weaknesses. It is a preventive way to reduce cyber threats. 

\subsection{Malware evasion insights}

\begin{figure}[!ht]
    \centering
    \includegraphics[width=0.8\textwidth]{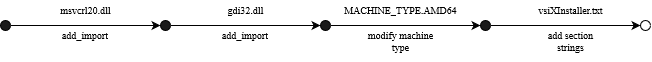}
    \caption{Extract of the vulnerability report where four steps are required to evade a malware}
    \label{fig:avex1}
\end{figure}

In this section, we describe a malware evasion on a commercial AV with the REINFORCE agent.
As we can see in figure \ref{fig:avex1}, the REINFORCE agent generates an undetectable malicious file in only four steps. The sequence is composed of the following actions \{8, 8, 0, 6\}. This figure also presents information extracted from the vulnerability report with the modifications and insertions made to the PE file. Two benign libraries are added with action 8  \{msvcrl20.dll, gdi32.dll\} and action 0 changes the machine type into AMD64. Then, the strings added with action 6 are extracted from a benign file named  \textquote{vsiXinstaller.txt}

\begin{figure}[!ht]
    \centering
    \includegraphics[width=0.7\textwidth]{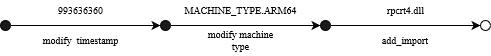}
    \caption{Extract of the vulnerability report where three steps are required to evade a malware}
    \label{fig:avex2}
\end{figure}

Figure \ref{fig:avex2} shows the details of the evasion of a second malicious file. Here, the agent takes only three steps to make the file undetectable. The required actions are \{12, 0, 8\}. Action 12 modifies the timestamps to  \textquote{993636360} and action 0 changes the machine type to ARM64. Finally, action 8 adds the library  \textquote{rpcrt4.dll}. 

We notice that the action \textquote{modify machine type} is used a few times in the test but has a zero cumulative score. However, in our two examples, the action is used as the second last action, so we can think it is significant in the evasion. This is a strength of REINFORCE compared to DQN, which learns action sequences and not only the optimal action. In the future, it will be interesting to isolate some actions, to really understand their impact on the evasion of the malware.

\section{Conclusion and future works}
\label{sec:conclusion}

\subsection{Conclusion}

\begin{comment}

In this paper, we have analysed different agents capable of evading several malware detection engines. Our prototype has the following advantages:

\begin{itemize}
    \item[$\bullet$] agents do not require any prior knowledge about the detection model or the malware to be evaded;
    
    %\item[$\bullet$] the evasion rate, once the agent is trained, is high thanks to the %complementarity of the REINFORCE and DQN algorithms;
    
    \item once the agent is trained, the evasion rate is high even on commercial AV thanks to the use of the REINFORCE algorithm.
    
    \item[$\bullet$] a useful vulnerability report is generated to help security analysts mitigate an evasion.
    
     \item[$\bullet$] its capability to generate new datasets of undetectable malware to re-train detection models.
\end{itemize}

\end{comment}

In this paper, we have analysed different agents capable of evading several malware detection engines. Our prototype has the advantages of not requiring any prior knowledge about the detection model and having a high evasion rate even on commercial AV, thanks to the use of REINFORCE algorithm. 
Our agents perform better than the random agent except on the commercial AV where the random agent is surprisingly better than DQN. This shows a limitation of DQN when we have only a hard label because DQN does not analyze the sequence of actions but only the last action that leads to an evasion. 
%On the other hand, the major concern of the random agent is the very high number of steps, which increases the chances of breaking the PE file. 
The conclusion of these results is that the DQN algorithm can be very effective in identifying the most impactful actions but has some strong limitations in black-box settings. On the contrary, the REINFORCE algorithm is more versatile as it gives very good results in all situations.
In addition to this, we are able to generate a useful vulnerability report to help security analysts mitigate an evasion. The prototype can also be used to generate new datasets of undetectable malware to re-train ML detection models. We believe that this work will improve malware detection tools in the future and strengthen antivirus software by providing analysts with vulnerability reports. We hope that this article will be able to contribute to the security of all in our fight against malware.

\subsection{Future works}
We aim to improve the technology presented here in several ways : 
\begin{itemize}
    \item optimizing the current agents, for example by training on particular malware families, like trojans or ransomware, because they are more widespread and more dangerous;
    \item implementing other RL algorithms type actor-critic (A2C, A3C \cite{mnih2016asynchronous}) to analyse their efficiency;
    \item in relation with defense, training, or re-training detection models with the new malicious files generated to determine if this increases their robustness and efficiency.
\end{itemize} 

\subsection{Code and database}
Despite the potential benefits of our work for strengthening AV defenses, we do not provide any open-access to our prototype since it may be used by malicious actors. The prototype is highly effective on every commercial AVs we have evaluated it on, so we believe that the risks of sharing our code exceed the possible benefits. However, we are open to dialogue with AV vendors to share our findings and to envision collaboration in the context of malware detection enhancement.

\subsection{Acknowledgments}

We want to thank the following colleagues Daniel Juteau, Philippe Calvet, Sok-yen Loui and Adam Ouorou. We are very grateful to BODMAS team for their valuable dataset \cite{bodmas}.

\cleardoublepage

\bibliographystyle{unsrt}
\bibliography{ref}

\vspace{12pt}

\end{document}